\documentclass{article}
\usepackage{spconf,amsmath,graphicx,hyperref}

\usepackage{booktabs} 
\usepackage{multirow} 

\title{From Contrast to Commonality: Audio Commonality Captioning for Enhanced Audio-Text Cross-modal Understanding in Multimodal LLMs
}
\name{Yuhang Jia\textsuperscript{\rm 1}, Xu Zhang\textsuperscript{\rm 1}, Yujie Guo\textsuperscript{\rm 1}, Yang Chen\textsuperscript{\rm 1}, Shiwan Zhao\textsuperscript{\rm 1}$^{*}$\thanks{$^{*}$ Corresponding author. This work has been supported by the National Key R\&D Program of China through grant 2022ZD0116307 and NSF China (Grant No.62271270).}}
\address{\textsuperscript{\rm 1}TMCC, College of Computer Science, Nankai University, Tianjin, China\\
Email: 2013628@mail.nankai.edu.cn, zhaosw@gmail.com
}

\begin{document}
\ninept
\maketitle
\begin{abstract}

Audio Captioning (AC) plays a pivotal role in enhancing audio-text cross-modal understanding during the pretraining and finetuning of Multimodal LLMs (MLLMs). To strengthen this alignment, recent works propose Audio Difference Captioning (ADC), which takes multiple audio inputs and encourages the model to describe their differences, thereby promoting fine-grained discrimination. However, despite its effectiveness, ADC introduces a semantic gap between input audios—often rich in diverse events—and the brief, difference-focused short caption. This deviation from AC-style task causes a mismatch with the pretraining objective, leading to catastrophic forgetting. To address this, we propose Audio Commonality Captioning (ACC), a comparably challenging but gentler alternative that guides the model to capture shared semantics across audio clips rather than detailed differences. Experiments show that ACC not only improves audio-text understanding on captioning benchmarks but also better preserves general capabilities across diverse speech and music tasks, confirming its ability to enable more robust cross-modal understanding and achieve a better balance between generalization and task-specific performance in MLLMs.

\end{abstract}

\vspace{-1mm}
\section{Introduction}
\vspace{-2mm}
Audio Captioning (AC) stands out as a primary task for audio–text cross-modal understanding, aiming to generate semantically aligned textual descriptions that accurately capture both the content and the style of audio signals. Benefiting from the availability of large-scale paired audio-text datasets such as AudioCaps, Clotho, and Audioset\cite{kim2019audiocaps, gemmeke2017audio, drossos2020clotho, mei2024wavcaps}, traditional AC models have achieved promising performance on audio captioning benchmarks\cite{mei2021encoder, gong2021ast, mei2022diverse, xu2024towards}. With the advent of multimodal large language models (MLLMs), audio captioning has gradually become a key pretraining and finetuning task to enhance models’ capabilities in audio analysis and understanding\cite{chu2023qwen, chu2024qwen2, xu2025qwen2, ding2025kimi}, as well as a crucial benchmark for evaluating these models’ performance in audio-text cross-modal understanding\cite{yang2024air, wang2024audiobench}.

\begin{figure}[t!]
    \centering
    \includegraphics[width=0.49\textwidth]{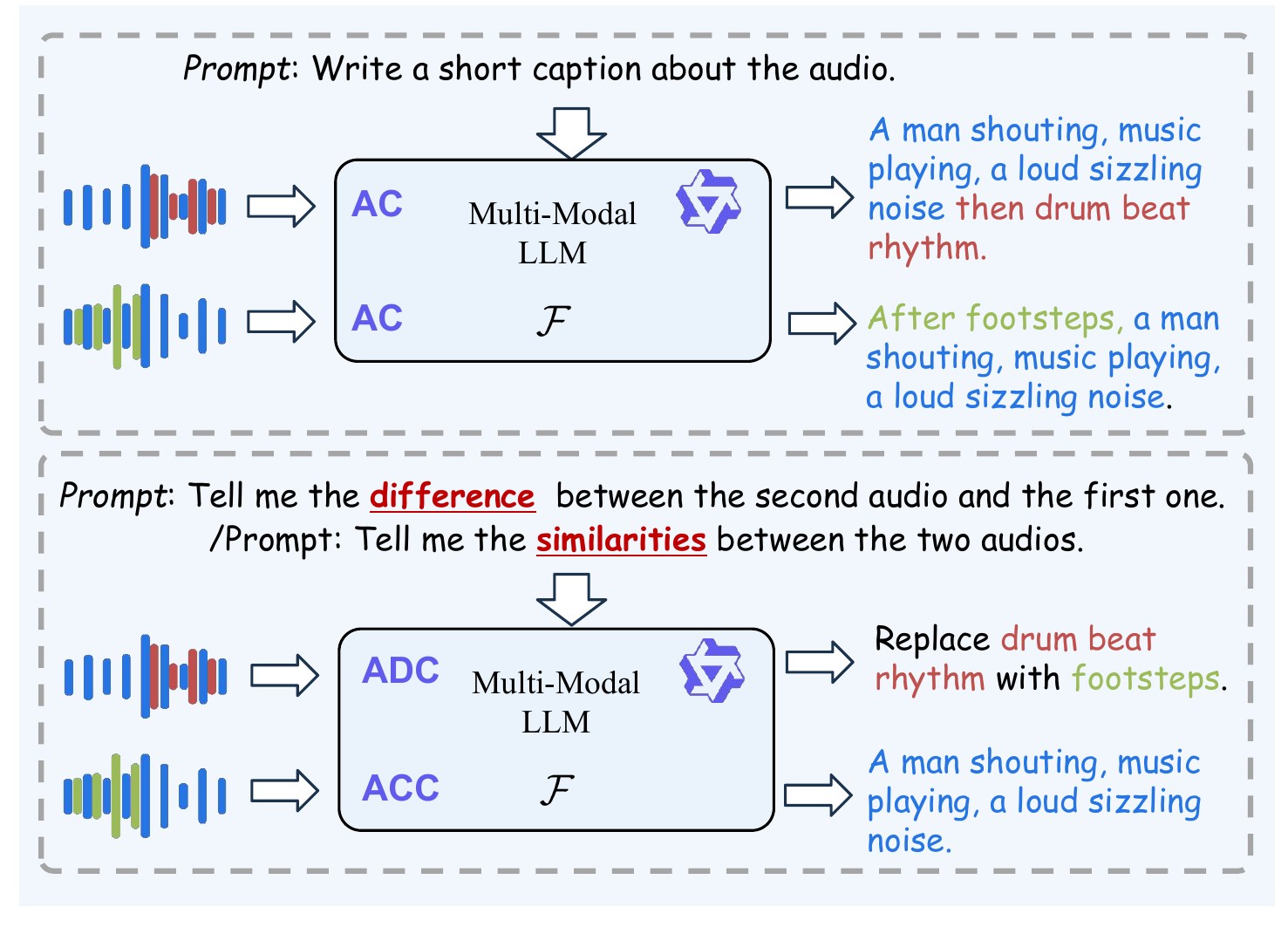}
    \vspace{-8mm}
    \caption{Comparison of Input, Prompt, and Objective among Audio Captioning, Audio Difference and Commonality Captioning.\vspace{-5mm}}
    \label{fig:model}
\end{figure}

Beyond traditional Audio Captioning (AC), Audio Difference Captioning (ADC) has emerged to address a key limitation—namely, the tendency of models to generate overly similar captions for acoustically related audio clips, thereby failing to capture fine-grained semantic distinctions \cite{wijngaard2025audio}. Instead of treating each audio clip in isolation, ADC explicitly incorporates contrastive information between paired inputs, encouraging models to highlight salient and discriminative elements that differentiate the clips. This contrastive formulation has demonstrated notable improvements on captioning benchmarks, underscoring its potential for advancing fine-grained and discriminative audio understanding.

Early progress in this direction includes Audio-Change Captioning \cite{kawaguchi2023audio}, which introduced difference captioning to the audio domain by training models to describe meaningful acoustic changes between paired inputs, thereby enhancing sound event interpretation. Building upon this foundation, subsequent works have developed more refined ADC frameworks \cite{kim2022exploring, komatsu2024audio, takeuchi2023audio}. For instance, PairMix \cite{kim2022exploring} applies multi-audio mixing as a data augmentation strategy, significantly improving downstream captioning and retrieval performance. Komatsu et al. \cite{komatsu2024audio} propose audio difference learning, where auxiliary reference audio guides the model to capture contextual differences, leading to gains in captioning quality. Meanwhile, Takeuchi et al. \cite{takeuchi2023audio} formally define the ADC task, introducing a transformer-based architecture with similarity–discrepancy disentanglement and releasing the AudioDiffCaps as a benchmark for difference captioning. Together, these efforts underscore the importance of difference-aware modeling for advancing fine-grained audio understanding.

However, despite its effectiveness in enabling difference-telling capabilities, ADC introduces a significant semantic shift between the rich input audios and the shorter output caption (see Figure~\ref{fig:model}). This divergence from AC-style descriptions creates a large gap from the pretraining objective, often leading to catastrophic forgetting during finetuning of MLLMs. To address this issue, we propose a novel captioning paradigm—\textbf{Audio Commonality Captioning (ACC)}—which encourages models to focus on the predominant shared semantics between audio clips rather than their detailed differences (see Figure~\ref{fig:model}). This gentler yet comparably challenging objective maintains consistency with the original AC formulation and fosters more effective audio-text cross-modal understanding. Our main contributions are as follows:

1) \textbf{Proposal of Audio Commonality Captioning }: We introduce ACC as a novel captioning paradigm that emphasizes shared semantic content between paired audio inputs, providing a gentler and more consistent training objective aligned with AC pretraining.

2) \textbf{Empirical validation on primary captioning benchmarks}: Experiments demonstrate that, in the context of MLLMs, ACC significantly improves audio-text alignment and caption quality compared to both conventional AC and ADC methods.

3) \textbf{Improved generalization on downstream tasks}: We show that ACC better preserves model's original capabilities across diverse speech and music-related tasks—including vocal sound classification (VSC), speech emotion recognition (SER), musical instrument classification (MIC) and music genre classification (MGC)—striking a better balance between generalization and task-specific performance.

\vspace{-4mm}
\section{Method}
\vspace{-2mm}
\subsection{Construction of the ADC and ACC Dataset}
\vspace{-2mm}
To construct paired audio–text datasets suitable for tasks such as ADC and ACC, prior work has primarily relied on simple concatenation or audio mixing strategies \cite{kim2022exploring, komatsu2024audio, takeuchi2023audio}. For example, some methods synthesize paired audio clips by artificially combining foreground event sounds with background contexts drawn from environmental sound datasets. A similar requirement for paired audio–text data arises in another class of tasks—Audio Editing—where each data sample consists of a pair of audio clips (before and after editing), an editing instruction, and textual descriptions corresponding to both versions. Such data supports the training of models that learn to modify audio content in accordance with textual instructions.

Audit \cite{wang2023audit} is the first framework to construct large-scale audio editing datasets by combining audio mixing with logical composition, supporting a wide range of editing operations including add, delete, replace, inpainting, and super-resolution. Its key advantage lies in scalability: given a base audio segment $A$ and two standalone event clips $B$ and $C$, numerous meaningful editing pairs can be generated through simple mixing and permutation. For example, mixing $A$ with $B$ or $C$ yields $A+B$ and $A+C$, which can be arranged into six distinct editing pairs: two \textbf{Add} examples ($A \rightarrow A+B$, $A \rightarrow A+C$); two \textbf{Delete} examples ($A+B \rightarrow A$, $A+C \rightarrow A$); and two \textbf{Replace} examples ($A+B \rightarrow A+C$, $A+C \rightarrow A+B$). This method enables the generation of a large number of training examples from a relatively small set of base audio clips and event sounds, making it a scalable solution for constructing editing datasets. 

In our work, we adopt the mixing strategy proposed in \cite{wang2023audit, jia2025towards} to construct training datasets for both the ADC and ACC tasks, thereby ensuring high-quality, instruction-aligned paired audio–text data for model training. Specifically, we use the AudioCaps dataset as the primary source of base audio segments, which serve as the $A$ component in each mixing operation. The single-event audio clips from AuditEval\cite{jia2025towards} are employed as the $B$ and $C$ event audios to be mixed with $A$. Notably, AuditEval provides a diverse collection of sound events spanning eight broad categories—NaturalSounds, Transportation, IndoorActivities, OutdoorActivities, Animals, Speech, Music, and HumanSounds—covering 100 distinct sound types and 400 unique clips. By combining these sources, we are able to synthesize a total of 148,500 audio editing pairs. Table~\ref{tab:statistics} summarizes the quality and quantity of the mixed audio. Compared to the original unedited audio, the mixed audio exhibits a slight decrease in text adherence and overall quality, yet still maintains a sufficiently high standard to provide reliable support for subsequent experiments.

We then employ AuditEval’s description generation pipeline to create both pre- and post-edit captions, together with the corresponding editing instructions. Based on this data, we construct two complementary datasets. For ADC, each sample consists of an original–edited audio pair as input, with the associated editing instruction (e.g., “add a burst of bird song”) serving as the target difference caption. For ACC, we leverage the same audio pairs but instead derive captions that describe the content shared between the pre- and post-edit audio clips, following the strategy detailed below:

1) For \textbf{Add} operations, we use the original (pre-edit) audio caption as the commonality, since it describes the content shared by both audio clips before the new element is added.
    
2) For \textbf{Delete} operations, we take the caption of the edited (post-edit) audio as the commonality, as it reflects the remaining content after removing the specified segment, which is still present in both.

3) For \textbf{Replace} operations, we align the pre- and post-edit captions at the word level and extract the longest overlapping phrase or content segment. This overlap, which remains unchanged before and after editing, is used as the commonality.

\begin{table}[h!]
\centering
\vspace{-6mm}
\caption{Quality and quantity statistics of the mixed clips.}
\label{tab:statistics}
\renewcommand{\arraystretch}{0.9}
\begin{tabular}{ccccc}
\toprule
\multirow{2}{*}{Audio Clips} & \multicolumn{2}{c}{Quality} & \multicolumn{2}{c}{Quantity}\\
\cmidrule(lr){2-3} \cmidrule(lr){4-5} 
 & Clap↑ & IS↑ & Utt. & Dur.(hrs) \\
\midrule
A (from audiocaps) & 51.81 & 13.12& 24750& 68.75\\
A mixed B & 49.49& 10.95& 24750& 68.75\\
A mixed C & 49.63& 11.86& 24750& 68.75\\
\bottomrule
\vspace{-7mm}
\end{tabular}
\normalsize
\end{table}

\begin{figure}[t!]
    \centering
    \includegraphics[width=0.50\textwidth]{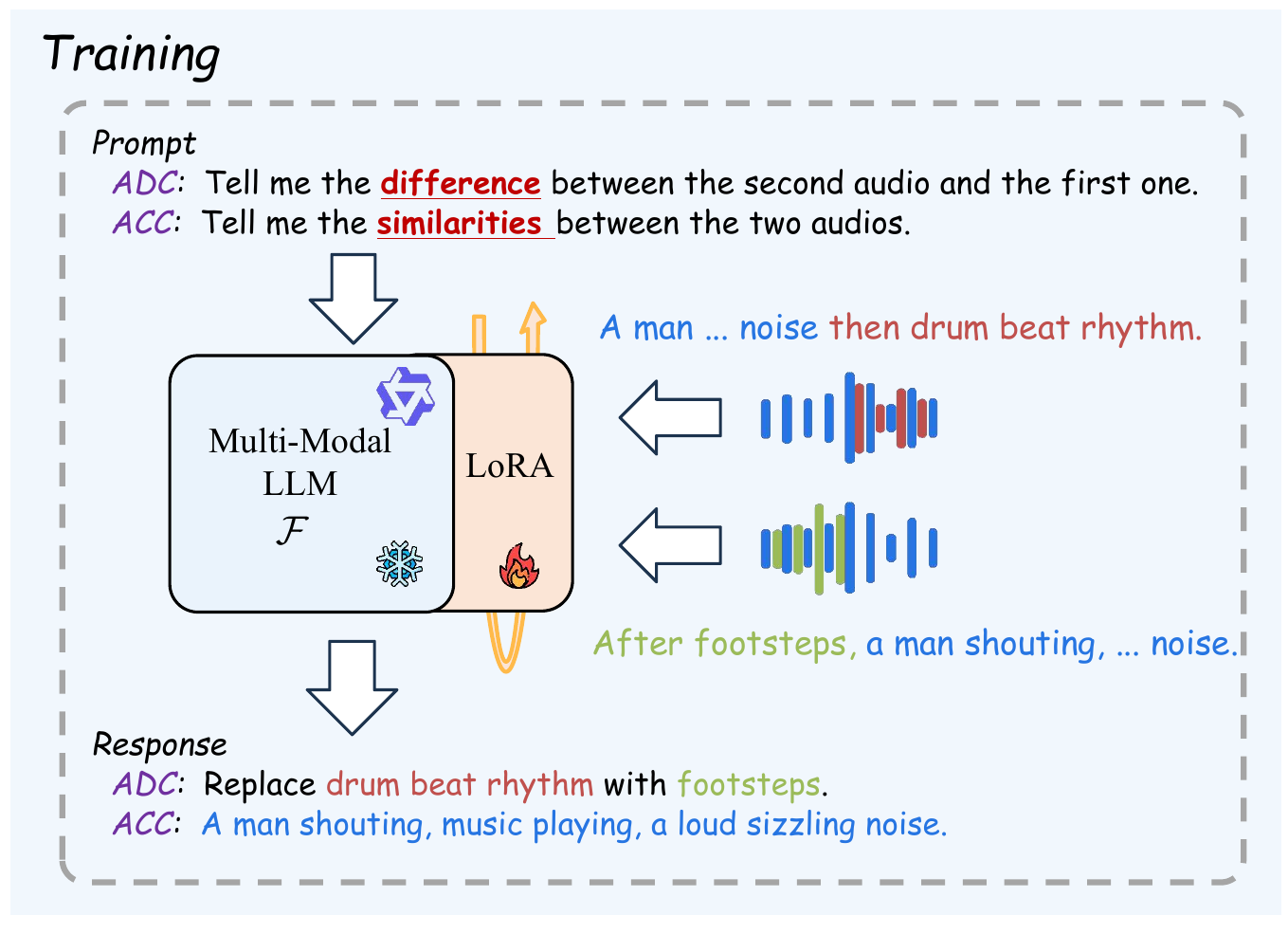}
    \vspace{-9mm}
    \caption{Illustration of fine-tuning the multi-modal LLMs on various audio-caption tasks using a lightweight LoRA-based adaptation.\vspace{-5mm}}
    \label{fig:Train}
\end{figure}

\begin{table*}[h!]
\centering
\caption{Audio captioning performance comparison on AudioCaps and Clotho benchmarks. Higher scores indicate better performance.}
\label{tab:audiocaption_performance}
\renewcommand{\arraystretch}{0.9}
\begin{tabular}{ccccccccccc}
\toprule
\multirow{3}{*}{Captioning\_Tasks} & \multicolumn{10}{c}{AudioCaps} \\
\cmidrule(lr){2-11}
& Bleu\_1 & Bleu\_2 & Bleu\_3 & Bleu\_4 & Fense & Spice & Spider & Cider\_d & Meteor & Rouge\_l \\
\midrule
Qwen2-Audio + AC & 0.3896 & 0.2648 & 0.1836 & 0.1301 & 0.7247 & 0.2742 & 0.7948 & 1.3155 & 0.2040 & 0.3827 \\
Qwen2-Audio + ADC & 0.0825 & 0.0309 & 0.0144 & 0.0051 & 0.3442 & \textbf{0.4480} & 0.0688 & 0.0927 & 0.0422 & 0.0949 \\
Qwen2-Audio + ACC (ours) & \textbf{0.4382} & \textbf{0.3422} & \textbf{0.2832} & \textbf{0.2452} & \textbf{0.7538} & 0.3701 & \textbf{1.4200} & \textbf{2.4699} & \textbf{0.2398} & \textbf{0.4557} \\
\midrule
Qwen2-Audio + AC + ADC & 0.3361 & 0.2195 & 0.1450 & 0.0976 & 0.7033 & 0.2392 & 0.6434 & 1.0475 & 0.1815 & 0.3478 \\
Qwen2-Audio + AC + ACC (ours)  & \textbf{0.4991} & \textbf{0.4014} & \textbf{0.3376} & \textbf{0.2941} & \textbf{0.7840} & \textbf{0.4111} & \textbf{1.6904} & \textbf{2.9698} & \textbf{0.2730} & \textbf{0.5078} \\
\midrule
\midrule
\multirow{3}{*}{Captioning\_Tasks} & \multicolumn{10}{c}{Clotho} \\
\cmidrule(lr){2-11}
& Bleu\_1 & Bleu\_2 & Bleu\_3 & Bleu\_4 & Fense & Spice & Spider & Cider\_d & Meteor & Rouge\_l \\
\midrule
Qwen2-Audio + AC & 0.1762 & 0.0894 & 0.0506 & 0.0295 & \textbf{0.5922} & 0.1529 & \textbf{0.2831} & \textbf{0.4132} & 0.1005 & 0.1762 \\
Qwen2-Audio + ADC & 0.0798 & 0.0262 & 0.0118 & 0.0053 & 0.3861 & 0.0566 & 0.0754 & 0.0943 & 0.0442 & 0.0798 \\
Qwen2-Audio + ACC (ours) & \textbf{0.1999} & \textbf{0.1039} & \textbf{0.0585} & \textbf{0.0338} & 0.5804 & \textbf{0.1534} & 0.2826 & 0.4118 & \textbf{0.1032} & \textbf{0.1999} \\
\midrule
Qwen2-Audio + AC + ADC & 0.1659 & 0.0842 & 0.0471 & 0.0271 & 0.5787 & 0.1442 & 0.2640 & 0.3839 & 0.0965 & 0.2078 \\
Qwen2-Audio + AC + ACC (ours)  & \textbf{0.2050} & \textbf{0.1033} & \textbf{0.0567} & \textbf{0.0314} & \textbf{0.5871} & \textbf{0.1460} & \textbf{0.2747} & \textbf{0.4034} & \textbf{0.1035} & \textbf{0.2138} \\
\bottomrule
\vspace{-7mm}
\end{tabular}
\end{table*}

\vspace{-2mm}
\subsection{Multimodal LLM Fine-tuning}
\vspace{-1mm}
To validate the effectiveness of Audio Commonality Captioning over Audio Difference Captioning in enhancing audio-text understanding in MLLMs, we fine-tune Qwen2-Audio-7B\footnote{\url{https://huggingface.co/Qwen/Qwen2-Audio-7B}}, an advanced and openly accessible audio-text MLLMs, on our constructed dataset. Qwen2-Audio extends the Qwen2 series by integrating a high-capacity audio encoder with a text-based decoder, and incorporates the Qwen-7B large language model to decode textual tokens. This architecture enables unified reasoning across acoustic and linguistic modalities. Moreover, Qwen2-Audio natively supports multi-audio inputs, making it particularly well-suited for our tasks like ACC and ADC that rely on jointly modeling multiple audio clips.

We adopt an instruction-tuning paradigm and apply LoRA-based parameter-efficient fine-tuning (see Figure~\ref{fig:Train}). During training, the model is conditioned on both audio inputs and Difference/ Commonality-telling instructions, and supervised to generate semantically appropriate, instruction-aligned responses tailored to each task. Through LoRA-based fine-tuning on ADC or ACC, the model is guided to capture fine-grained semantic structures essential for improved captioning, which are crucial for achieving stronger cross-modal alignment. This training strategy is expected to bridge the representational gap between audio and text, thereby enhancing the model’s robustness in scenarios where audio inputs contain subtle variations and demand precise semantic discrimination.

\vspace{-2mm}
\section{Experiments}
\vspace{-1mm}
\begin{figure*}[t!]
    \centering
    \includegraphics[width=0.99\textwidth]{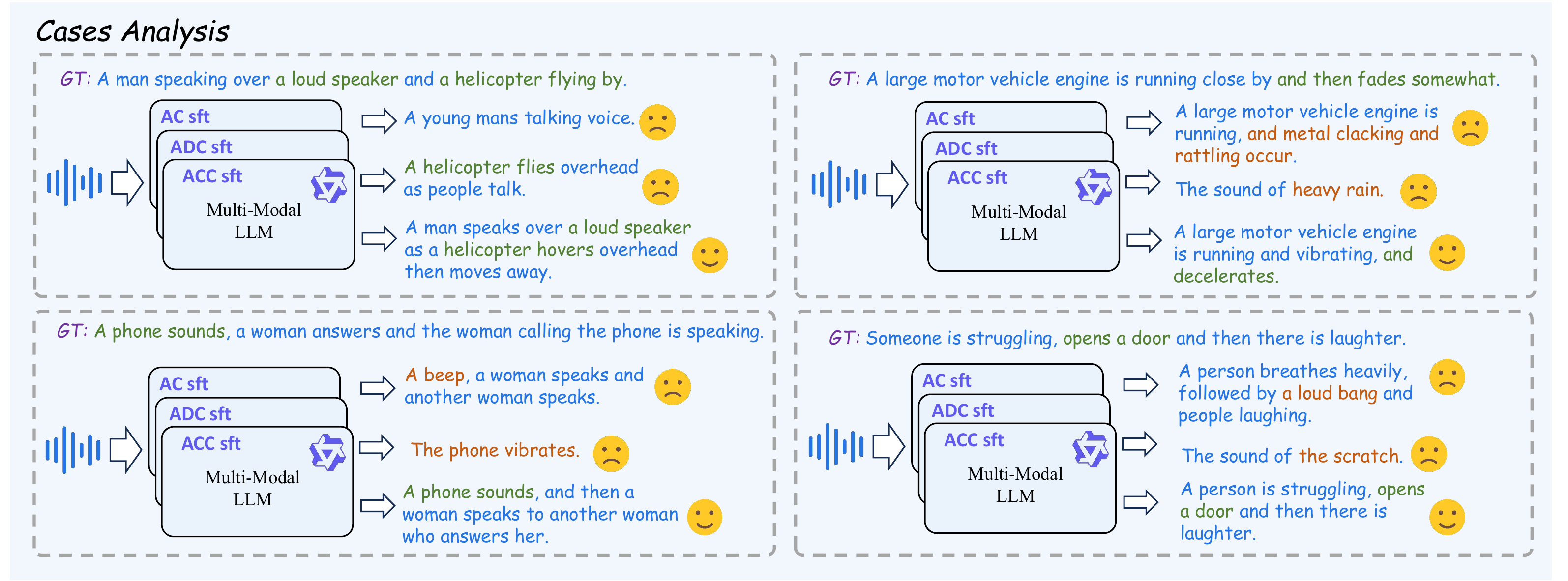}
    \vspace{-4mm}
    \caption{Comparison of fine-grained audio event understanding cases among baseline AC, ADC, and our proposed ACC. }\vspace{-5mm}
    \label{fig:cases}
\end{figure*}

\subsection{Dataset and Metrics}
\vspace{-2mm}
In this study, we investigate the relative impact of direct Audio Captioning (AC) fine-tuning, Audio Difference Captioning (ADC), and Audio Commonality Captioning (ACC) on enhancing the audio-text cross-modal understanding of MLLMs. We adopt the audio captioning task as the primary evaluation benchmark, as it directly reflects the model’s ability to achieve semantic alignment between audio and text before and after fine-tuning. For evaluation, we utilize two widely used datasets AudioCaps~\cite{kim2019audiocaps} and Clotho~\cite{drossos2020clotho}. It is worth noting that AudioCaps is partially incorporated into the construction of our fine-tuning datasets for AC, ADC, and ACC through sampling and mixing, which may introduce some overlap between pretraining and evaluation data. Nevertheless, within the context of MLLMs, such strategies remain valuable, as they offer meaningful empirical insights and practical benefits, particularly when they yield substantial performance improvements during large-scale adaptation. In contrast, the Clotho dataset serves as a complementary benchmark without any data overlap. For AC evaluation, we report a comprehensive suite of widely adopted captioning metrics, including BLEU (BLEU-1 to BLEU-4), FENSE, METEOR, ROUGE-L, CIDEr-D, SPICE, and SPIDEr, using the official evaluation toolkit\footnote{\url{https://github.com/Labbeti/aac-metrics}}.

To further examine the impact of fine-tuning on ADC and ACC with respect to the model’s generalization across downstream tasks, we evaluate performance on a diverse set of speech- and music-related tasks, including vocal sound classification (VSC), speech emotion recognition (SER), musical instrument classification (MIC), and music genre classification (MGC). For each task, we adopt widely used benchmark datasets: VocalSound~\cite{gong_vocalsound} for VSC, IEMOCAP~\cite{busso2008iemocap} for SER, NSynth~\cite{engel2017neural} for MIC, and GTZAN~\cite{sturm2013gtzan} for MGC. Accuracy is reported as the primary evaluation metric.
\vspace{-3mm}
\subsection{Training Details}
\vspace{-2mm}
All experiments were conducted on a single NVIDIA GeForce RTX 4090 GPU. We fine-tuned Qwen2-Audio-7B-Instruct using LoRA (rank=8, $\alpha$=32, dropout=0.05) implemented via ms- swift\footnote{\url{https://github.com/modelscope/ms-swift}}
. Training employed AdamW (lr=$1\mathrm{e}{-4}$, weight decay=0.1) with cosine scheduling, batch size 1, and gradient accumulation of 16. 
\vspace{-3mm}
\subsection{Results and Analysis}
\vspace{-1mm}

As shown in Table~\ref{tab:audiocaption_performance}, we compare the impact of three fine-tuning tasks—AC, ADC, and ACC—on the performance of audio caption  across the AudioCaps and Clotho datasets. On AudioCaps, the ACC task substantially outperforms both traditional AC and ADC, achieving state-of-the-art results on nearly all metrics except SPICE. In particular, ACC yields large gains in CIDEr-D (2.4699 vs. 1.3155 for AC), SPIDEr (1.4200 vs. 0.7948), and BLEU1–4 scores, underscoring its superior ability to strengthen cross-modal alignment and produce captions that are semantically closer to human annotations.
In contrast, ADC does not enhance the model’s fine-grained cross-modal understanding in the context of MLLMs and even causes notable degradation on several metrics compared to standard AC.

On Clotho, ACC again outperforms competing approaches, achieving state-of-the-art results on most metrics, including BLEU1 to4, SPICE, METEOR, and ROUGE-L. This demonstrates that the improvements brought by ACC generalize beyond the training distribution and genuinely enhance the model’s audio understanding. Although AC slightly surpasses ACC on FENSE and SPIDEr, the differences are marginal, and ACC overall demonstrates more balanced and robust performance. Similarly, ADC continues to underperform in the MLLM fine-tuning setting. These results show that fine-tuning with ACC enables MLLMs to internalize more fine-grained cross-modal semantic structures, resulting in higher-quality captions while avoiding catastrophic forgetting during fine-tuning.

To further illustrate these advantages, Figure~\ref{fig:Train} presents several examples of fine-grained audio event understanding across AC, ADC, and our proposed ACC on the audio captioning task. The results clearly demonstrate ACC’s superiority in both robustness and fine-grained comprehension. While the AC model captures general events, it fails to correctly identify subtle sounds, such as “Sound decelerates,” “Phone rings,” or “Door opens.” In contrast, the ADC model shows poor robustness and suffers from catastrophic forgetting, often generating single-event or vague descriptions like “Sound of the scratch” or “Phone vibrates.” ACC, however, accurately captures detailed events and maintains high semantic fidelity, producing captions that closely match the ground truth. This case study further highlights ACC’s ability to enhance the model’s understanding of nuanced acoustic events, particularly in fine-grained scenarios.

\begin{table}[t!]
\centering
\caption{Performance on speech- and music-related tasks. The original model serves as a reference to assess how well different fine-tuning strategies preserve performance on other tasks.}
\label{tab:performance_comparison}
\renewcommand{\arraystretch}{1.0}
\begin{tabular}{ccccc}
\toprule
\multirow{3}{*}{Tasks} & VSC & SER & MIC & MGC \\
\cmidrule(lr){2-5}
& VocalSd & IEMOCAP & NSynth & GTZAN \\
\midrule
Qwen2-Audio & 93.61\% & 62.85\% & 65.50\% & 70.67\% \\
\midrule
+ AC & \textbf{93.65\%} & \textbf{65.03\%} & \underline{59.86}\% & \underline{70.97\%} \\
+ ADC & 81.29\% & 58.50\% & 59.33\% & 67.57\% \\
+ ACC(ours) & \underline{93.00\%} & \underline{61.72\%} & \textbf{61.87\%} & \textbf{72.07\%} \\
\bottomrule
\vspace{-9mm}
\end{tabular}
\setlength{\tabcolsep}{6pt}
\normalsize
\end{table}

Table~\ref{tab:performance_comparison} presents the performance comparison on different peech and music-related tasks, illustrating how these tasks affect the model’s generalization ability across various downstream applications. Notably, the model fine-tuned on the standard AC task achieves the best performance on two speech-related tasks—VSC and SER—even surpassing the original model. While the model fine-tuned on the ACC task also shows competitive results, significantly outperforming the one trained on ADC task. This suggests that ACC is still effective at preserving model' original capabilities related to voice and speech, offering an entirely acceptable balance between generalization and the task-specific performance.

For the two music-related tasks—MIC and MGC—the ACC task demonstrates a superior ability to preserve the model’s original performance. Specifically, for the MIC task, the model fine-tuned with ACC maintains the highest performance compared to those fine-tuned with AC or ADC. Remarkably, for the MGC task, the ACC-tuned model even improves upon the original model’s accuracy (from 70.67\% to 72.07\%). We hypothesize that this improvement stems from the construction of the ACC dataset, where the selected single-audio events may contain descriptions related to music, thereby contributing to enhanced music-related capabilities. In contrast, the model fine-tuned on ADC exhibits inferior generalization across all four tasks, suggesting its limitations in fine-tuning scenarios for unified multi-task MLLMs.

To mitigate the potential influence of data overlap from pretraining and better reflect a realistic training pipeline, where a multimodal model is first pretrained on AC task before being further adapted to finer-grained tasks such as ADC or ACC, we fine-tuned the models on both the AudioCaps corpus (for AC) and one of the finer-grained datasets (ADC or ACC), as shown in Table \ref{tab:audiocaption_performance} bottom. This setup enables the model to jointly learn general audio semantics and finer-grained distinctions. 
From Table \ref{tab:audiocaption_performance}, we observe consistently stronger performance when applying ACC for fine-grained audio-text optimization alongside the traditional AC task, compared to the AC+ADC setting. This pattern holds across both the AudioCaps and Clotho datasets. On AudioCaps, ACC achieves substantially higher scores in CIDEr-D (2.9698 \textit{vs.}1.0475), SPIDEr (1.6904 \textit{vs.}0.6434), and BLEU metrics (e.g., BLEU-4: 0.2941 \textit{vs.}0.0976). On Clotho, although the differences are more modest, ACC still surpasses AC+ADC on key metrics such as BLEU-4 (0.0314 vs. 0.0271) and CIDEr-D (0.4034 vs. 0.3839). These results demonstrate that incorporating ACC as a fine-grained refinement task alongside AC consistently leads to more effective audio-text alignment and richer semantic understanding across diverse datasets, establishing ACC as a superior fine-tuning strategy for enhancing audio-text cross modal understanding in the context of multimodal large language models.

\vspace{-3mm}
\section{Conclusion}
\vspace{-2mm}
We propose Audio Commonality Captioning, a novel training paradigm for audio captioning that maintains fine-grained supervision while providing a gentler and more consistent training objective aligned with AC pretraining. Through comprehensive experiments on both in-domain and out-of-domain benchmarks, we show that ACC consistently outperforms standard Audio Captioning and Audio Difference Captioning in enhancing audio-text cross-modal understanding. Further evaluations demonstrate that ACC better preserves the model’s general capabilities across various speech and music-related tasks, achieving a more favorable balance between generalization and task-specific performance. These findings establish ACC as a effective and versatile training objective for advancing audio-text understanding in multimodal large language models.

\vfill\pagebreak
\small
\bibliographystyle{IEEEbib}
\bibliography{strings,refs}

\end{document}